\documentclass[
 superscriptaddress,
 reprint,
 amsmath,amssymb,
 aps,
]{revtex4-1}

\usepackage{graphicx}
\usepackage{dcolumn}
\usepackage{bm}
\renewcommand{\vec}[1]{\mathbf{#1}}
\usepackage{amsmath}
\usepackage{amsfonts}
\usepackage{amssymb}
\usepackage{dsfont}
\usepackage{tikz}
\usepackage{wrapfig}
\usepackage{braket}
\usepackage[caption=false]{subfig}
\usepackage{float}
\usepackage{placeins}
\usepackage{epstopdf}
\usepackage{hyperref}
\usepackage[percent]{overpic}
\usepackage{xcolor}
\usepackage{nicefrac}

\usepackage{graphicx}
\graphicspath{C:/Users/mjgba/Documents/codes/anisotropic/}

\newcommand{\cre}{\hat b^\dag}
\newcommand{\an}{\hat b}
\newcommand{\ex}[1]{\langle #1 \rangle}
\newcommand{\NN}[1]{< #1 >}

\newcommand{\n}{\hat n}

\newcommand{\jump}{\par \noindent}

\begin{document}

\title{Supersolid phases of ultracold bosons trapped in optical lattices dressed with Rydberg $p$-states}

\author{Mathieu Barbier}
\email{barbier@itp.uni-frankfurt.de}
\affiliation{Institut f\"ur Theoretische Physik, Goethe-Universit\"at, 60438 Frankfurt/Main, Germany}
\author{Henrik L\"utjeharms}
\affiliation{Institut f\"ur Theoretische Physik, Goethe-Universit\"at, 60438 Frankfurt/Main, Germany}
\author{Walter Hofstetter}
\affiliation{Institut f\"ur Theoretische Physik, Goethe-Universit\"at, 60438 Frankfurt/Main, Germany}

\date{\today}

\begin{abstract}
Engineering quantum phases with spontaneously broken symmetries is a major goal of research in different fields. Trapped ultracold Rydberg-excited atoms in optical lattices are a promising platform for realizing quantum phases with broken lattice translational symmetry since they are interacting over distances larger than the lattice constant. Although numerous theoretical works on trapped Rydberg-excited gases have predicted such phases, in particular density wave or supersolid phases, their experimental observation proves to be difficult due to challenges such as scattering processes and the limited experimentally achievable coupling strength. Most of these previous studies have focused on isotropically interacting gases dressed with Rydberg $s$-states, while the effect of anisotropic interactions due to Rydberg-excited $p$-states in trapped quantum gases remains much less investigated. Additionally, it was shown that the excitation scheme used to excite Rydberg $p$-states possesses advantages regarding achievable coupling strengths and limitation of scattering processes compared to its $s$-state counterpart, which makes the investigation of Rydberg $p$-state dressed quantum gases even more interesting. In the present work, we study the extended, two-component Bose-Hubbard model, realized with a bosonic quantum gas with Rydberg-excited $p$-states trapped in an optical lattice, within Gutzwiller mean-field theory. We compute the ground state phase diagram and investigate its different regimes. By comparison to the phase diagram of the isotropic case, we find the anisotropic interaction to be more advantageous for the observation of supersolid phases.
\end{abstract}

\maketitle


\section*{Introduction}
Due to their exaggerated properties, atoms coupled to Rydberg states are prominent candidates for a manifold of experimental setups, such as quantum computation \cite{QuantumComputing1,QuantumComputing2,QuantumComputing3} and quantum simulation of lattice spin models \cite{SpinLattice1,SpinLattice2,RydbergTweezer} in optical lattices and tweezers. Longer-range interactions of atoms in optical lattices can be induced by the van-der-Waals interaction, which is introduced through coherent coupling of trapped atoms to Rydberg-excited states. Since the length scale of the long-range interaction is typically larger than the lattice constant, novel quantum phases such as lattice supersolids - phases with simultaneously broken lattice translational and $U(1)$ symmetry \cite{RydbergSupersolidsI,RydbergSupersolidsII,
RydbergSupersolidsIII,RydbergSupersolidsIV} - appear to be within reach.\\
Although theoretical studies of interacting atomic lattice gases dressed with Rydberg $s$-states found parameter regimes for which supersolids should be experimentally observable \cite{RydbergPhasediagram1,RydbergPhasediagram2,RydbergPhasediagram3},
different obstacles have so far made the experimental realization of supersolids challenging. The lifetimes of these systems are limited through scattering processes and were shown to be much smaller than the typical single particle lifetime due to collective loss processes \cite{SpinLattice1,AvalancheII,Atomloss2}. Furthermore an additional AC Stark shift arising from the coupling laser impedes possible coherent tunneling of particles and shifts the necessary hopping amplitude to unreasonably high values \cite{ACStark1,ACStark2}.\\
One possibility to address these challenges is through coherent coupling to Rydberg $p$-states. While one substantial difference between Rydberg $s$-states and $p$-states is the geometry of the induced long-range interaction, which is anisotropic between Rydberg $p$-states atoms \cite{LevelShifts,Anisotropic,AnisotropicMagnetic}, a crucial benefit lies in the corresponding coupling scheme. Coupling a hyperfine ground state to a Rydberg $s$-state requires coupling schemes with at least one intermediate state, due to the total angular momentum conservation according to the dipole selection rule, and thus several coupling lasers, which leads to smaller achievable coupling strengths and additional losses \cite{SinglePhotonI,SinglePhotonII}. In contrast, the single-photon coupling scheme allows for larger achievable coupling and ratio between interaction strength and scattering rates, which is promising for the observation of supersolid quantum phases.\\
\noindent
In this work we investigate a bosonic quantum gas with coherent coupling to a Rydberg $p$-state trapped in a two-dimensional optical lattice.
In the first section, we introduce the extended, two-component Bose-Hubbard model, explain its Rydberg physics related features and discuss its tunability and limitations.
We then introduce Gutzwiller mean-field theory and its application to the previously defined Hamiltonian. We discuss the validity of the mean-field approximations, which are used to decouple non-local terms and to derive a set of effective single-site Hamiltonians. We conclude that within the parameter regime of the subsequent calculations the applied mean-field approximations are valid and yield qualitatively good results.\\
\noindent
In the second section we discuss and compare the various ground state phase diagrams obtained for the extended, two-component Bose Hubbard model. In addition to the Mott-insulating and superfluid phases, we find spatially modulated phases, e.g. density wave and supersolid phases, which arise from the long-range interaction. Furthermore, we obtain a devil's staircase of density wave phases and find a two-stage melting of density wave phases. Finally, we compare the ground state phase diagram for non-tilted anisotropic, tilted anisotropic and isotropic interaction, and find a wider parameter regime for the supersolid phases in the case of non-tilted anisotropic interaction.\\

\section{System and method}
\begin{figure}[t]
\includegraphics[width = 1\linewidth, trim = {120 0 0 0}, clip]{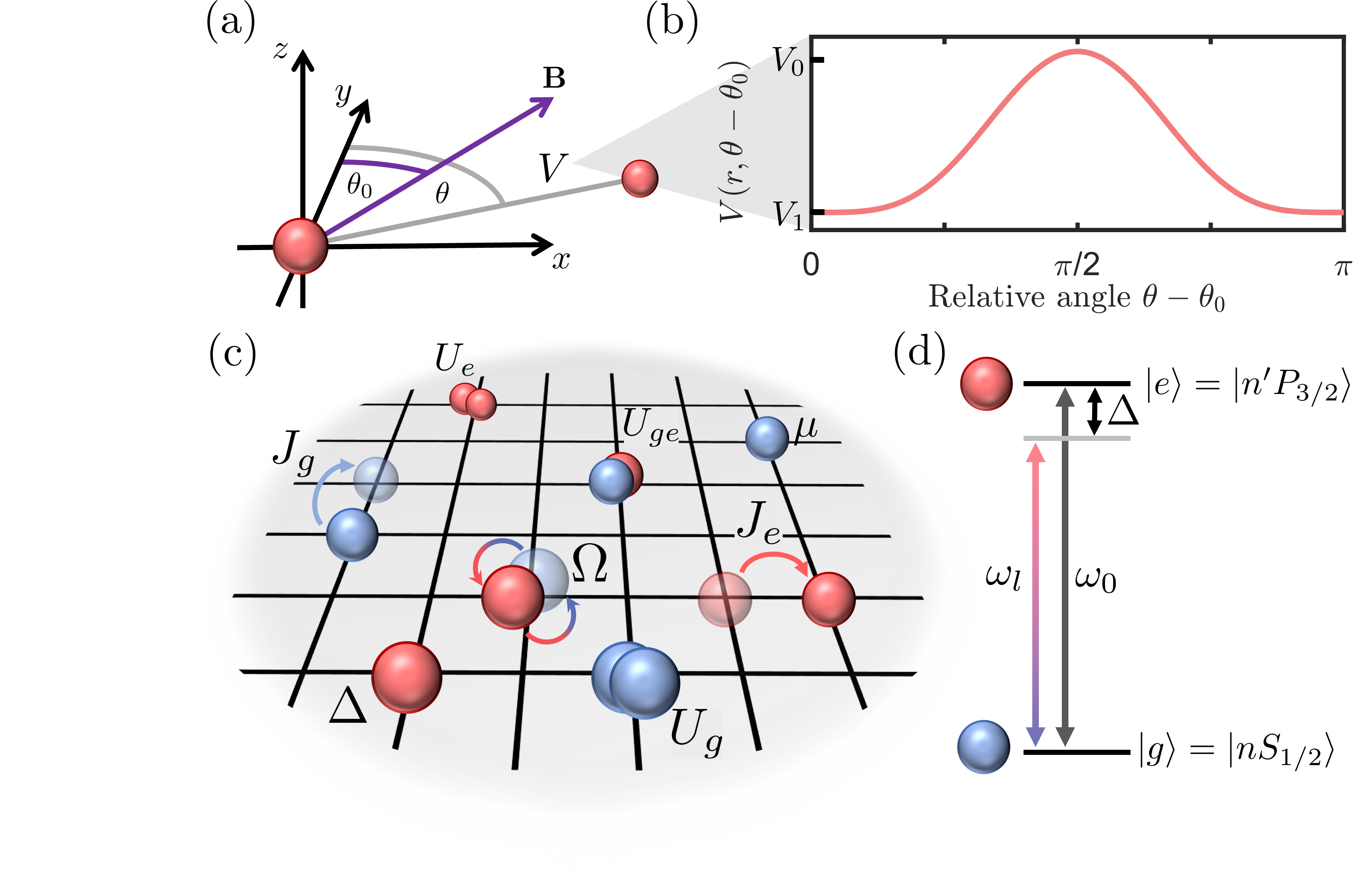}
\caption{(a) Orientation of the magnetic field given by the angle $\theta_0$. The magnetic field $\vec{B}$ serves as a reference axis for the anisotropic interaction. (b) Interaction $V(r,\theta-\theta_0)$ for the full range of relative angle $\theta - \theta_0$, which is $V_0$ at $\theta-\theta_0 = n\pi$ and approximately $V_1$ at $\theta-\theta_0 = (m + 1/2)\pi$ with $n,m \in \mathbb{N}$ for $V_0 \gg V_1$. (c) Schematic of the processes corresponding to the different terms in the Hamiltonian \eqref{eq:Hamiltfull}. The particles are subject to on-site potentials given by $\mu$ and $\Delta$, on-site interactions with strength $U_g$, $U_{ge}$ and $U_e$, hopping with strength $J_g$ and $J_e$ and inter-state conversion with Rabi frequency $\Omega$. (d) Single-photon coupling scheme characterized by the Rabi coupling $\Omega$, the transition frequency $\omega_0$, the laser frequency $\omega_l$ and the detuning $\Delta$.}
\label{fig:scheme}
\end{figure}

\begin{figure}
\includegraphics[width = 1\linewidth, trim = {0 0 0 0}, clip]{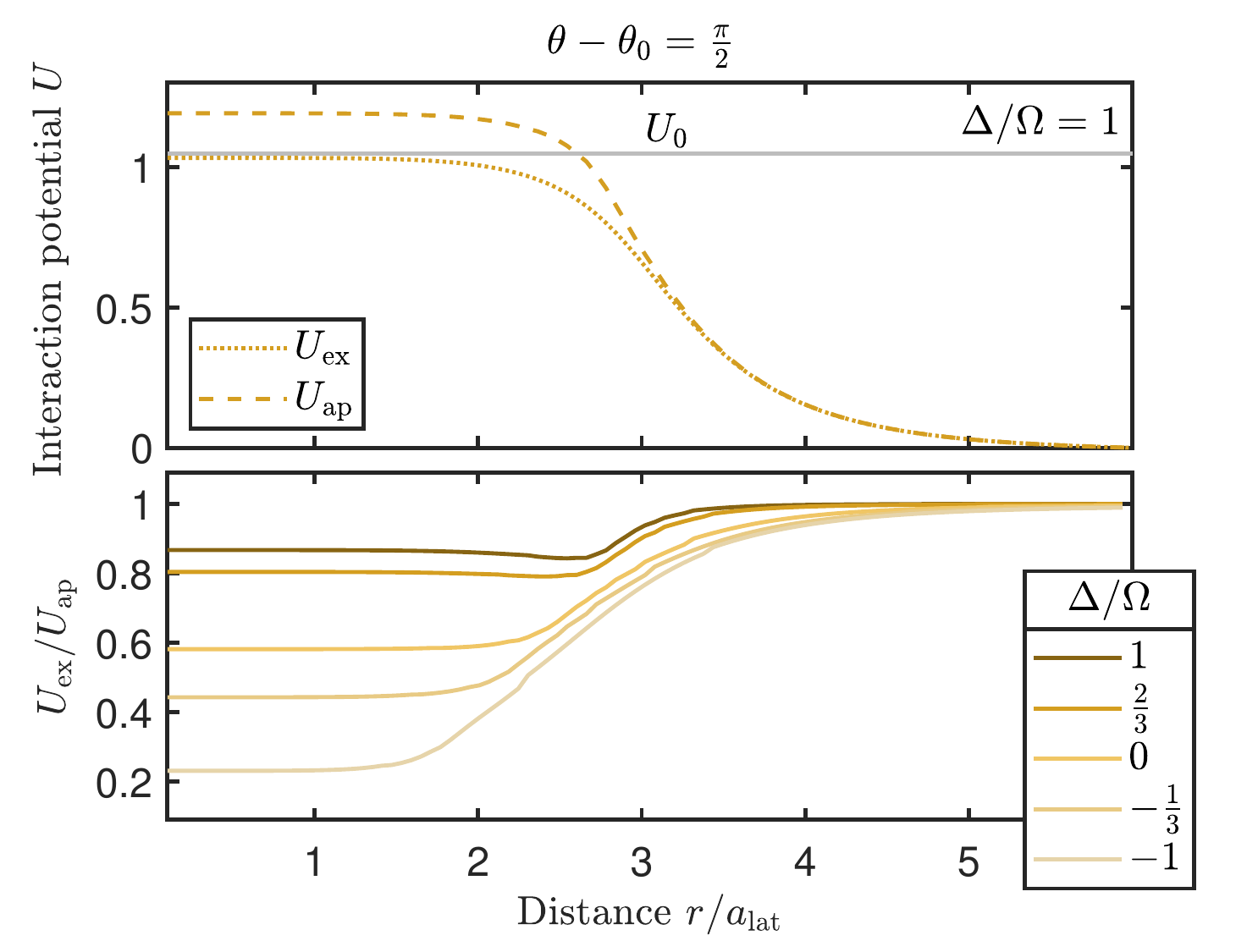}
\caption{Upper figure: Exact soft-core potential $U_\mathrm{ex}$ of the two-body Hamiltonian (dotted line) and the soft-core potential $U_\mathrm{ap}$ within the Hartree-approximation (dashed line) for detuning $\Delta/\Omega = 1$ and $\theta - \theta_0 = \pi/2$. Lower figure: Ratio $U_\mathrm{ap}/U_\mathrm{ex}$ between the exact and the approximated soft-core potentials for various detunings. While for positive detuning both soft-core potentials are in good agreement, we obtain a pronounced difference for small absolute or negative values of the detuning.}
\label{fig:Hartree}
\end{figure}

In this work, we consider a bosonic quantum gas trapped in a two-dimensional optical lattice, e.g. $^{87}$Rb \cite{RbP} or $^{133}$Cs \cite{CsP,CsP2}, coherently coupled to a Rydberg $p$-state through an external coupling laser. The single photon coupling scheme in the single-atom basis is characterized by the Rabi coupling $\Omega$ and the detuning $\Delta = \omega_0 - \omega_l$, the difference between the transition frequency $\omega_0$ and the laser frequency $\omega_l$ (see FIG. \ref{fig:scheme}). We study the corresponding extended, two-component Bose Hubbard model. The corresponding Hamiltonian can be written as
\begin{equation}
\hat{H} = \hat{H}_g + \hat{H}_e + \hat{H}_{ge} \label{eq:Hamiltfull}
\end{equation}
with the electronic ground (excited) state Hamiltonian $\hat{H}_{g(e)}$ and the inter-state Hamiltonian $\hat{H}_{ge}$. The atoms in the electronic ground state are best described by the single-component Bose-Hubbard model
\begin{equation}
\hat{H}_g = - J_g \sum_{<ij>} (\hat{b}^g_{i})^\dag \hat{b}^g_j + \frac{1}{2}U_g\sum_i \hat{n}^g_{i}(\hat{n}^g_{i}-1) - \mu \sum_i \hat{n}^g_i \label{eq:Hamiltground}
\end{equation}
with the ground state hopping amplitude $J_g$, the on-site interaction between ground state atoms with strength $U_g$ and the chemical potential $\mu$. The Hamiltonian of the electronic excited state is
\begin{equation}
\begin{split}
\hat{H}_e = &- J_e \sum_{<ij>} (\hat{b}^e_i)^\dag \hat{b}^e_j + \frac{1}{2}U_e\sum_i \hat{n}^e_i(\hat{n}^e_i-1) - \mu \sum_i \hat{n}^e_i\\
&-\Delta \sum_i \hat{n}^e_i + \sum_i \sum_{j \neq i} V_{ij} \hat{n}^e_i\hat{n}^e_j, \label{eq:Hamiltexcited}
\end{split}
\end{equation}
with the detuning $\Delta$ and the anisotropic van-der-Waals interaction
\begin{equation}
V_{ij} = \frac{V_0 \mathrm{sin}^4(\theta - \theta_0) + V_1}{|\vec{r_i}-\vec{r_j}|^6},
\end{equation}
whose strength is characterized by $V_0$ and $V_1$ \cite{Anisotropic}. The reference axis of the anisotropic interaction is given by the reference angle $\theta_0$, and is experimentally realized and tunable through an external magnetic field $\vec{B}$ (see FIG. \ref{fig:scheme}) \cite{AnisotropicMagnetic,SinglePhotonI,MagneticField}.\\
The inter-state Hamiltonian
\begin{equation}
\hat{H}_{ge} = U_{ge} \sum_i \hat{n}^g_i \hat{n}^e_i + \frac{\Omega}{2} \sum_i ((\hat{b}^g_i)^\dag \hat{b}^e_i + (\hat{b}^e_i)^\dag \hat{b}^g_i)
\end{equation}
describes an inter-state on-site interaction with strength $U_{ge}$ and the coherent driving between the ground state and the Rydberg-excited state with Rabi coupling $\Omega$.\\
The detuning and the coherent coupling term are obtained within the rotating wave approximation (RWA) \cite{RWAI,RWAII,RWAIII}. The RWA holds for $|\Delta| \ll |\omega_0 + \omega_l|$, which is typically satisfied in experiments, where values of the detuning are orders of magnitude smaller than both laser and transition frequencies.\\
\jump
As the dimension of the many-body Hilbert space grows exponentially with the number of atoms considered, a method for reducing the computational effort and making the calculation feasible is required. In this work, we treat the bosonic lattice system within Gutzwiller mean-field theory to explore the ground state phases of the model \cite{GutzwillerVII,GutzwillerVIII,GutzwillerTwoComponent}. Within this theory, the many-body ground state wave function $|\Psi\rangle$ factorizes as a product $| \Psi \rangle = \prod_i | \Psi \rangle_i$ over single-site wave functions $| \Psi \rangle_i$, such that we only need to self-consistently solve a set of single-site Schr\"odinger equations $\hat{H}_i | \Psi \rangle_i = E_0^i | \Psi \rangle_i$, coupled through mean-fields, in order to determine the many-body ground state. To this end, we need to split the Hamiltonian of the full system as a sum $\hat{H} = \sum_i \hat{H}_i$, where the single-site Hamiltonians $\hat{H}_i$ have yet to be explicitly defined. Although the local terms of the Hamiltonian $\hat{H}$ can be directly split into single-site parts, we need to apply mean-field approximations to decouple the non-local terms, i.e. the hopping and the longer-range interaction.\\
\noindent
Within Gutzwiller mean-field theory we decouple the non-local kinetic terms, while the Hartree-approximation is applied to the long-range interaction term. For both approximations, we expand an operator $\hat{O} \in \{\hat{b}_i^\nu,\hat{n}_i^e\}$ through its expectation value as $\hat{O} = \langle \hat{O} \rangle + \delta \hat{O}$, where $\delta \hat{O} = \hat{O} - \langle \hat{O} \rangle$ is the quantum fluctuation. The approximation is made by neglecting quantum fluctuations of second order or higher, i.e. $(\delta \hat{O})^n \approx 0$ for $n \geq 2$. We can hereby decouple the Hamiltonian $\hat{H}$ into the single-site Hamiltonians
\begin{equation}
\begin{split}
\hat{H}_i = &\sum_{\nu \in \{g,e\}} \Big( - J_\nu ((\hat{b}_i^\nu)^\dag \xi_i^\nu + \mathrm{h.c.}) + \frac{1}{2} U_\nu \hat{n}_i^\nu (\hat{n}_i^\nu - 1) \Big) \\
& - \mu \hat{n}_i  - \Delta\hat{n}_i^e + \hat{n}_i^e \eta_i + U_{ge} \hat{n}_i^g \hat{n}_i^e\\
&  + U_{ge} \hat{n}_i^g \hat{n}_i^e + \frac{\Omega}{2} ((\hat{b}^g_i)^\dag \hat{b}^e_i + \mathrm{h.c.})  + E_i^\text{off}
\label{eq:Hamiltapprox}
\end{split}
\end{equation}
with the occupation number operator $\n = \n^g_i + \n^e_i$, the mean-fields $\xi^\nu_i = \sum_{j \in \mathrm{NN}(i)} \langle \hat{b}_i^\nu \rangle$ with $\nu \in \{g,e\}$ and $\eta_i = 2 \sum_{j \neq i} V_{ij} \langle \hat{n}^e_j \rangle$, and the energy offset $E_i^\text{off}$ resulting from the decoupling (see Appendix A). Each single-site Hamiltonian and therefore its ground state depends on expectation values evaluated on other sites, which leads to self-consistency conditions coupling all single-site problems. Nevertheless, due to the reduced Hilbert space of each single-site problem numerical simulations become feasible even for large lattices.\\
The validity of Gutzwiller mean-field theory becomes arguable in the vicinity of phase transitions. Even though incorporating the neglected quantum fluctuations can shift phase boundaries, the mean-field approximation has been shown to qualitatively describe phase boundaries in ground state phase diagrams well \cite{ClusterLuhmann}.\\
In order to check the validity of the Hartree-approximation, we determine the interaction potential of two atoms at a distance $r$, each coupled to a Rydberg-excited $p$-state. It typically takes the form of a soft-core potential and is analytically given by
\begin{equation}
U(r) = \frac{U_0}{1+(r/r_c)^6}
\end{equation}
with the soft-core height $U_0$ and the characteristic range of the interaction $r_c$. The soft-core potential is numerically obtained through computation of the ground state energy of the corresponding two-body Hamiltonian. Since the Hamiltonian can be treated either exactly or within the Hartree-approximation, the calculation of the soft-core potential provides a valuable platform for benchmarking the Hartree-approximation (see Appendix B). We compute the exact soft-core potential $U_\text{ex}$ and the soft-core potential $U_\text{ap}$ within the Hartree-approximation, which we compare for varying detuning at $\theta - \theta_0 = \pi/2$ (see FIG. \ref{fig:Hartree}). In general, the soft-core potential obtained within the Hartree-approximation is larger than the exact soft-core potential. While the difference becomes less for increasing detuning $\Delta/\Omega$, negative detunings lead to a more significant mismatch. Furthermore, we do not find a change in the quality of the approximation upon variation of $\theta - \theta_0$ (see Appendix B).
In previous calculations on Rydberg-excited gases in optical lattices, ground state phase diagrams computed within Gutzwiller mean-field theory with Hartree-decoupling were in qualitatively good agreement with results obtained by other methods \cite{RydbergAndreas,RydbergPhasediagram2}. We therefore expect the Gutzwiller-mean field theory to deliver qualitatively, and in some regimes also quantitatively accurate results.\\
In the ground state phase diagram we expect different homogeneous and inhomogeneous phases with distinctive spatial distribution of local observables. In order to identify the different phases within the parameter space considered, we determine the spatial distribution of the condensate order parameter $\phi_\nu$ and the occupation number $n_\nu$ for both the electronic ground state ($\nu = g$) and the Rydberg-excited state ($\nu = e$), as well as their spatially averaged mean values $\bar{\phi_\nu}$ and $\bar{n}_\nu$ (see Appendix C).\\

\section{Ground state phase diagrams}
We calculate the ground state phase diagrams of the Hamiltonian \eqref{eq:Hamiltfull}, which possesses a number of parameters. While most of these parameters are highly tunable, some are set by intrinsic properties of Rydberg states and the interaction between them. First, we assume the relevant time scales of the excited state to be primarily given by the Rabi frequency of the coherent coupling and by the long-range interaction strength. The contribution of the excited state to the kinetic energy is expected to be small, which motivates us to assume a vanishing hopping rate, i.e. $J_e = 0$. Furthermore the on-site interaction between a Rydberg-excited atom and another atom in either state is set by the Quantum Zeno Effect \cite{ZenoEffect2,ZenoEffect4}. Atoms in the presence of two-body loss processes exhibit hard-core behavior in the limit of strong loss. For two atoms of which at least one is in a Rydberg-excited state the formation of molecular ions due to their large scattering cross sections has been observed \cite{ZenoEffect,ZenoEffect3}. Since these molecules are not trapped by an optical lattice and are therefore lost upon formation, we set the on-site interaction of the excited state to large values, i.e. $U_{ge}, U_e \rightarrow \infty $. The other parameters of the Hamiltonian can be either directly or indirectly tuned, for example through Feshbach resonances and varying the lattice depth \cite{SuperfluidMott,Wannier}. Unless mentioned otherwise, we set the reference axis to be the $y$-axis, i.e. $\theta_0 = 0$, which is induced experimentally through an external magnetic field oriented in the same direction and leads to maximum interaction strength along the $x$-axis. A finite value of the reference angle $\theta_0 \in (0,\pi/2)$ implies a reference axis different from the coordinate axis.\\
In the following, we present the results obtained via the single-site Hamiltonian \eqref{eq:Hamiltapprox}. Within periodic boundary conditions, we compute the many-body ground state of finite size systems with different superlattice unit cells for a fixed set of parameters (see Appendix C). Through comparison of their energies, we identify the true many-body ground state of the system as the state with the lowest energy. We perform the above-mentioned steps for every point in the considered parameter spaces and thus obtain the upcoming phase diagrams.

\begin{figure}[t]
\includegraphics[width = 1\linewidth, trim = {0 0 0 0}, clip]{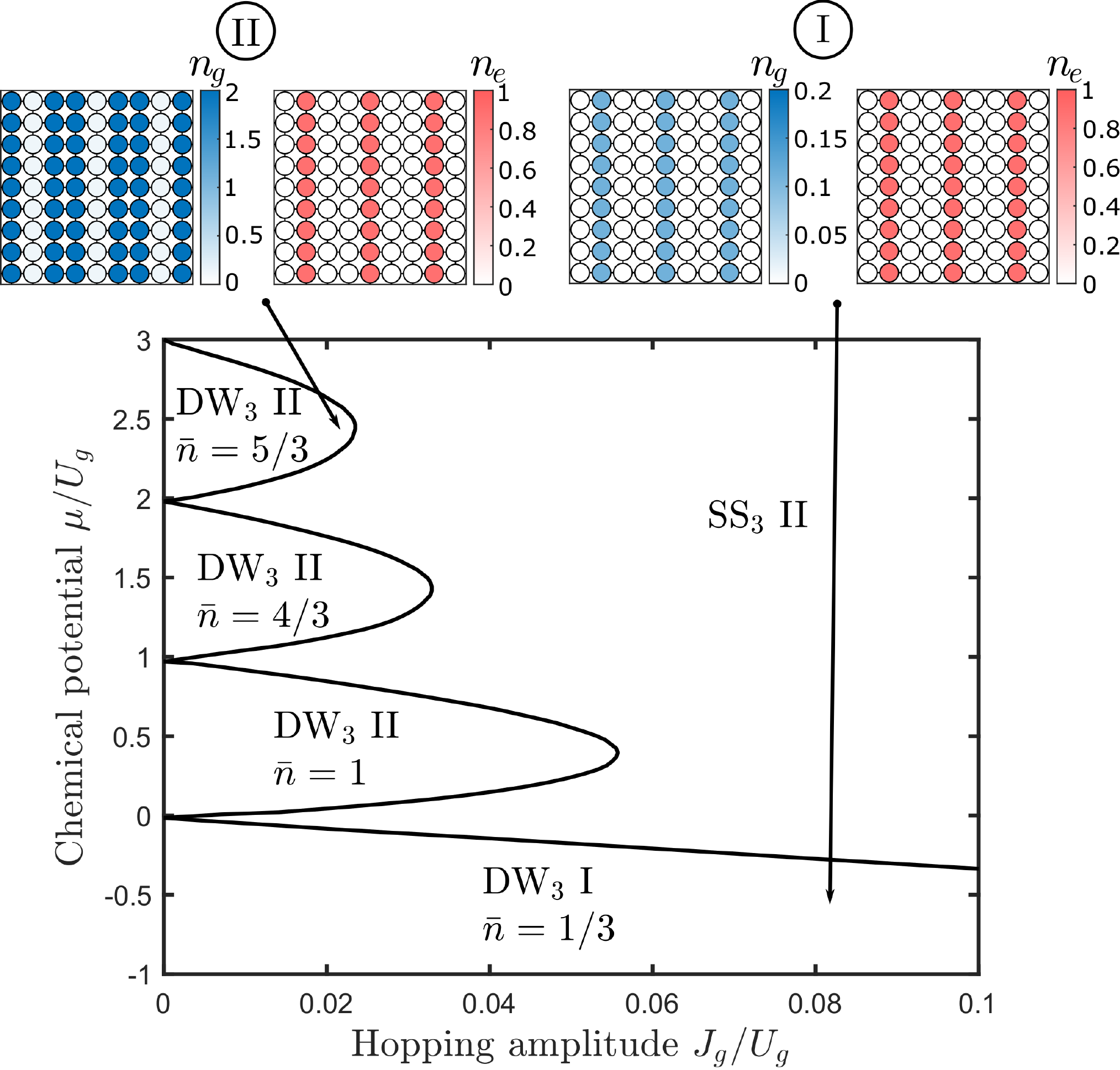}
\caption{$J_g - \mu$ phase diagram for anisotropic, long-range interaction at $U_g/\Omega = 0.1$, $U_{ge}/\Omega = 1$, $U_e/\Omega = 1000$, $\Delta/\Omega = 2$, $V_0/\Omega = 1000$, $V_1/\Omega = 1$, $\theta_0 = 0$ and $J_e/\Omega = 0$. The topology of the phase boundaries resembles the ones of the single species Bose-Hubbard model, although the vacuum, the Mott insulator and the superfluid regime have been replaced by two distinct density wave regimes (DW I and DW II) and a supersolid (SS II) regime. The indices denote the superlattice area $A_\text{SL}$ and the spatially modulated density is characterized by stripes, parallel to the reference axis in orientation, of excited-state atoms. The distance between stripes does not change here, since the parameters of the Hamiltonian relevant for the Rydberg state are kept fixed.}
\label{fig:mu_J}
\end{figure}

\subsection{$J_g - \mu$ phase diagram}
\begin{figure}[t]
\includegraphics[width = 1\linewidth, trim = {10 0 0 0}, clip]{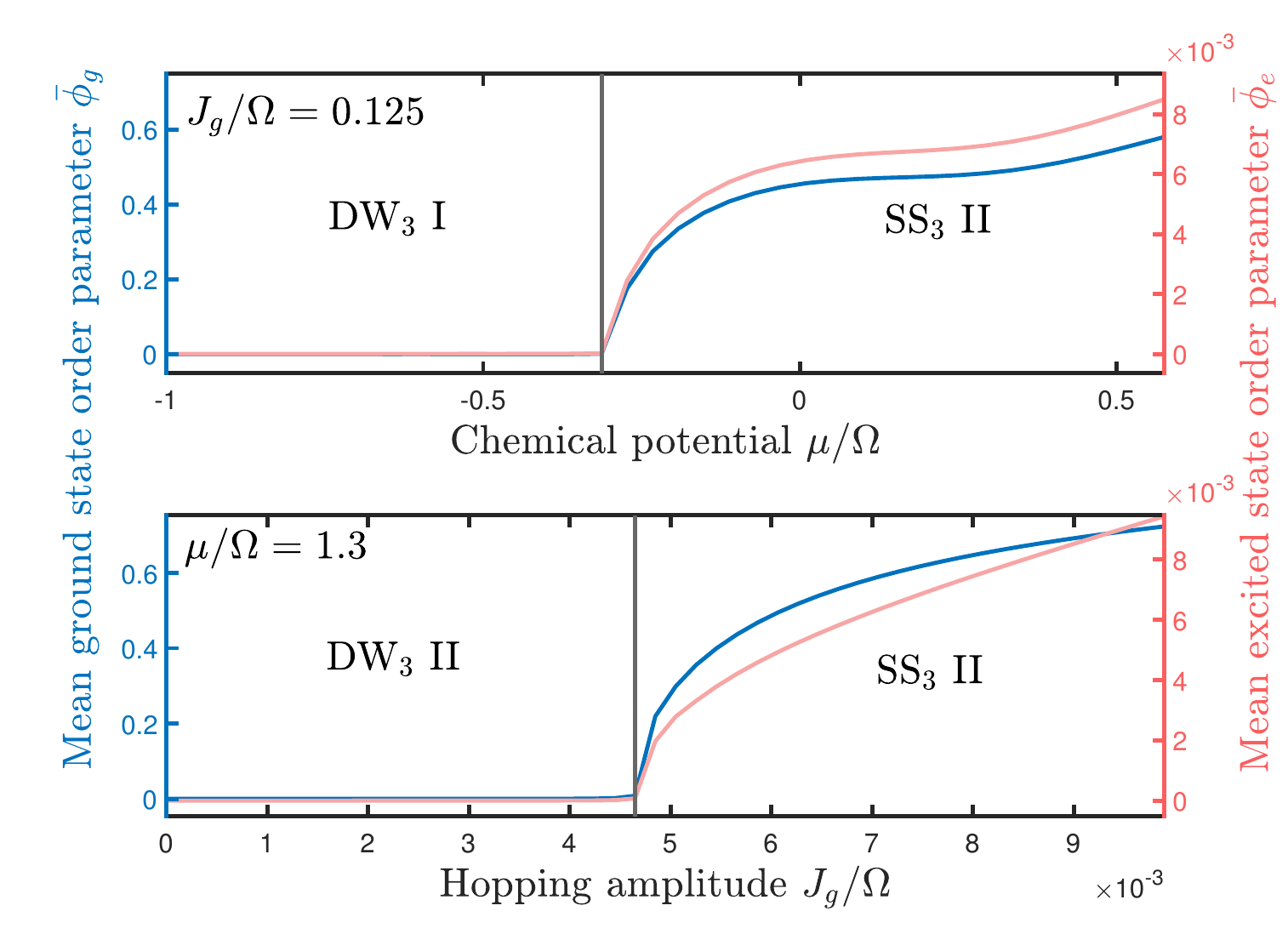}
\caption{Mean ground state order parameter $\bar \phi_g$ (dark blue line) and excited state order parameter $\bar \phi_e$ (light red line) at fixed hopping amplitude (upper diagram) and fixed chemical potential (lower diagram). The DW-SS phase transition is of second order upon parameter change, as the mean observables exhibit a kink at the transition.}
\label{fig:phase_transitions_BH}
\end{figure}

We first investigate the model by varying the (rescaled) hopping amplitude $J_g/\Omega$ of the ground state and chemical potential $\mu/\Omega$, and set the other parameters of the Hamiltonian to $U_g/\Omega = 0.1$, $U_{ge}/\Omega = 1$, $U_e/\Omega = 1000$, $\Delta/\Omega = 2$, $V_0/\Omega = 1000$, $V_1/\Omega = 1$, $\theta_0 = 0$ and $J_e/\Omega = 0$. We obtain an ground state phase diagram with phase boundaries that resemble the ones of the phase diagram of the single-species Bose-Hubbard model (see FIG. \ref{fig:mu_J}). The Mott-insulating (MI), superfluid (SF) and a vacuum (vac) regimes are replaced, however, by different density wave (DW) phases - insulating phases with broken lattice translational symmetry - and a supersolid (SS) regime with finite condensate order parameter and broken translational symmetry, both induced by the long-range interaction.
As expected, the anisotropic interaction leads to striped phases. Since in FIG. \ref{fig:mu_J} the detuning and the long-range interaction strengths are kept fixed, the crystalline structure does not vary within the parameter space considered.
By taking a closer look at the spatial distribution of the observables within the different regimes (see FIG. \ref{fig:mu_J}), we identify two types of phases based on their density wave structure: Phases of the first type (denoted by I) have maximum occupation number of both species (ground state and Rydberg-excited state) at the same site, while the other sites are barely occupied. These phases usually emerge when the chemical potential $\mu/\Omega$ is negative, since in that case finite occupation of the electronic ground state only emerges due to the coherent coupling to the excited state. On the other hand, phases of the second type (denoted by II) have complementary occupation number of both species on each site and generally occur for positive chemical potential $\mu/\Omega$. While the chemical potential dictates the type of phase at zero hopping, we also see that finite hopping leads to a transition to a SS II regime, which extends even to negative chemical potentials for large enough hopping amplitude. Interestingly, the SS II phase retains the same type of density wave structure along the phase transition.\\
\noindent
We calculate the mean condensate order parameter of the ground and excited state upon varying $\mu$ or $J_g$, respectively, and identify second order (continuous) phase transitions (see FIG. \ref{fig:phase_transitions_BH}). The phase boundaries between the insulating regimes with $\bar{\phi}_\nu = 0$ and the regime of finite condensate order parameter, i.e. $\bar{\phi}_\nu \neq 0$, are of second order. This is similar to the MI-SF transition of the Bose-Hubbard model, which is also of second order for spinless bosons \cite{BH_secondorder}.\\
\subsection{$J_g - \Delta$ phase diagram}
We now investigate the parameter space spanned by the (rescaled)  hopping amplitude $J_g/\Omega$ of the ground state and detuning $\Delta/\Omega$, since these parameters are experimentally easily tunable. We set the other parameters of the Hamiltonian to $U_g/\Omega = 0.1$, $U_e/\Omega = 1$, $U_g/\Omega = 1000$, $V_0/\Omega = 1000$, $V_1/\Omega = 1$, $J_e/\Omega = 0$ and $\mu/\Omega = -0.25$. We first investigate the frozen limit ($J_g = 0$), since we expect a variation of the detuning to lead to a manifold of insulating, inhomogeneous phases arising from the long-range interaction, and then move to the itinerant regime ($J_g/\Omega > 0$), in which the ground state states are determined by the interplay between atomic motion and crystalline ordering.\\
\begin{figure}[t]
\includegraphics[width = 0.9\linewidth, trim = {0 0 20 0}, clip]{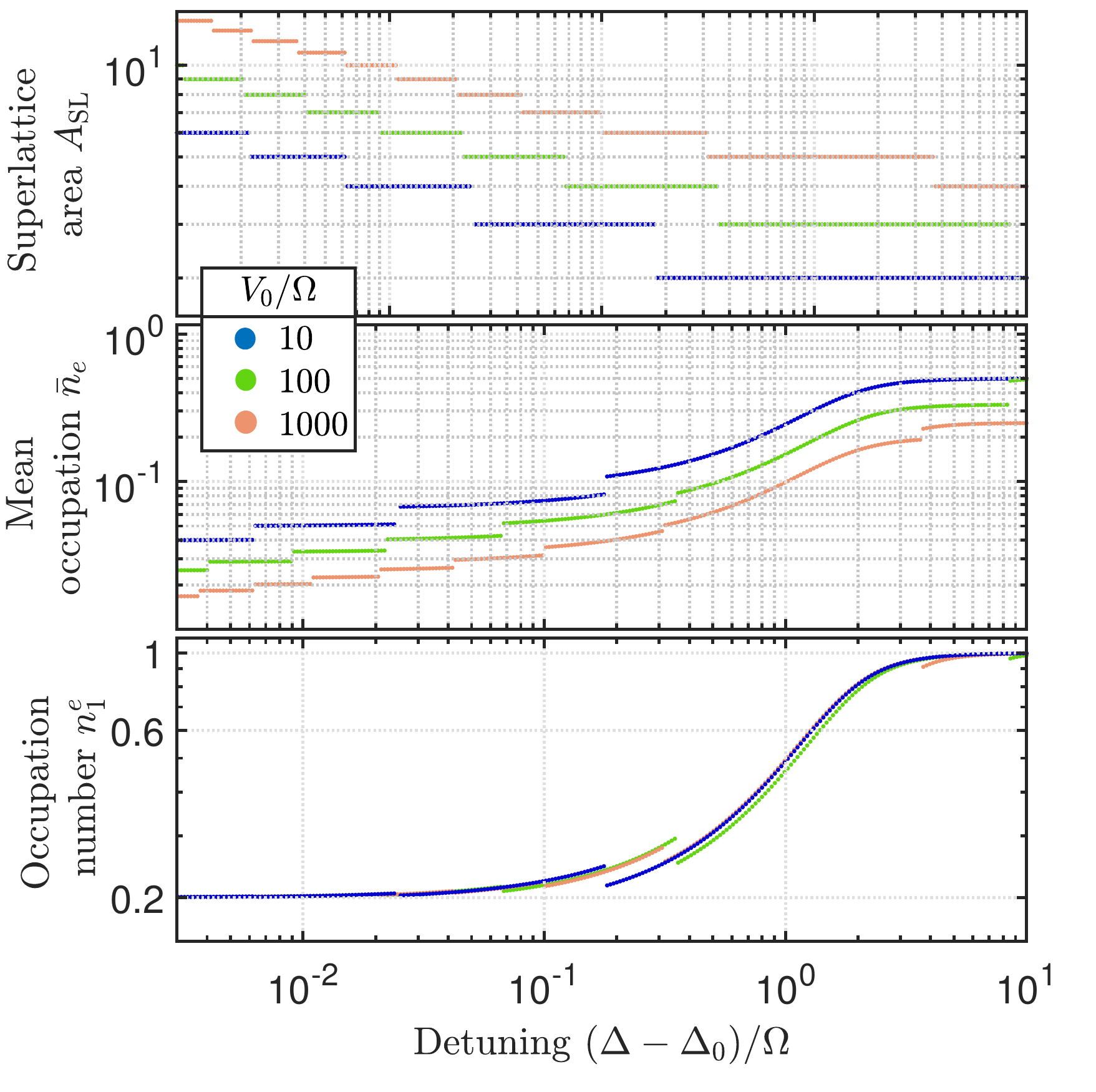}
\caption{The so-called devil's staircase (upper diagram), the mean occupation of the excited state $\bar{n}_e$ (middle diagram) and the excited state fraction $n_1^e$ of the occupied site (lower diagram) of crystalline density wave phases obtained by variation of the detuning $\Delta$ and various long-range interaction strengths $V_0/\Omega$ at $U_g/\Omega = 0.1$, $U_e/\Omega = 1$, $U_g/\Omega = 1000$, $V_1/\Omega = 1$, $J_e/\Omega = 0$ and $\mu/\Omega = -0.25$. While the crystalline structure and mean occupation number depend on the detuning and the long-range interaction strength, we find that the occupation number of the excited state of the lone occupied site within the superlattice unit cell only depends on the detuning.}
\label{fig:devil}
\end{figure}
\begin{figure}[t]
\includegraphics[width = 1\linewidth, trim = {10 0 10 0}, clip]{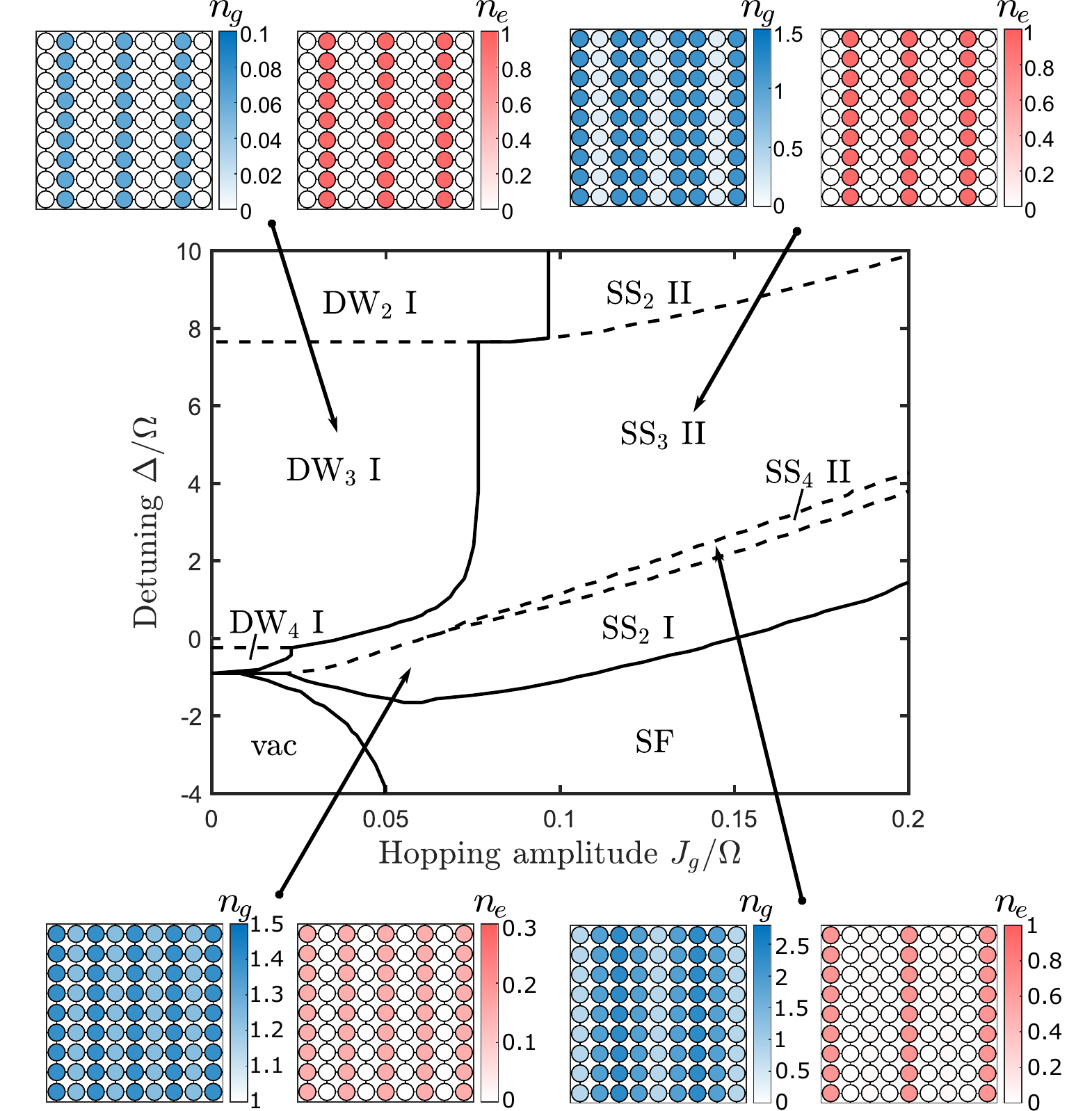}
\caption{$J_g - \Delta$ phase diagram for anisotropic, long-range interaction at $U_g/\Omega = 0.1$, $U_e/\Omega = 1$, $U_g/\Omega = 1000$, $V_0/\Omega = 1000$, $V_1/\Omega = 1$, $J_e/\Omega = 0$ and $\mu/\Omega = -0.25$. We find a density wave (DW), a supersolid (SS), a vacuum and a superfluid (SF) regime. The SS regime is composed of SS I and SS II phases with different spatial modulations. The subscript indices denote the superlattice area $A_\text{SL}$.}
\label{fig:anisotropic}
\end{figure}
\subsubsection{Frozen limit $J_g = 0$}
The frozen limit ($J_g = 0$) describes motionless Rydberg-excited atoms trapped in an optical lattice. Thus we expect a vanishing condensate order parameter and therefore crystalline structures of localized particles as the ground state states of the corresponding Hamiltonian \cite{Anisotropic}. Due to the anisotropic long-range interaction between atoms, we also expect striped phases oriented parallel to the reference axis of the anisotropic interaction, since the interaction is minimal along this direction. Since we expect the crystalline structures of the ground states to change through the variation of the detuning, we introduce the superlattice unit cell area $A_\text{SL}$ as a means to identify either sparse or dense packing of the excited-state atoms in the system (see Appendix C). For striped phases, the superlattice unit cell area quantifies the distance between stripes.\\
We find that varying the detuning $\Delta$ leads to a manifold of DW I phases with decreasing average occupation number $\bar{n}_e$ of the excited state as the detuning tends to a critical value $\Delta_0$ from positive values (see FIG. \ref{fig:devil}). We obtain no DW II phases due to the negative chemical potential $\mu/\Omega < 0$. Note that the superlattice unit cell area $A_\text{SL}$ increases as we decrease the detuning. Similar sequences of DW phases have also been found in the context of Rydberg-excited atoms with isotropic long-range interaction and are commonly referred to as devil's staircases \cite{Devil1,Devil2,Devil3}. Although the devil's staircase depends on interaction strength $V_0$ and Rabi coupling $\Omega$, we find that the occupation number $n^e_1$ of the single occupied site in the superlattice unit cell is approximately independent of these parameters and converges towards a finite value in the limit $\Delta \rightarrow \Delta_0$.\\
The critical detuning $\Delta_0$ marks the transition point from the density wave regime to the vacuum state. For a given chemical potential $\mu$ and Rabi coupling $\Omega$ the critical value is analytically given by $\mu = -(\Delta_0 + \sqrt{\Delta_0^2 + \Omega^2})/2$ \cite{RydbergPhasediagram2,RydbergPhasediagram3}. This results in our case in a critical detuning $\Delta_0/\Omega = -0.75$, which matches the value obtained in the numerical calculation.\\
\begin{figure}[t]
\includegraphics[width = 1\linewidth, trim = {0 0 0 0}, clip]{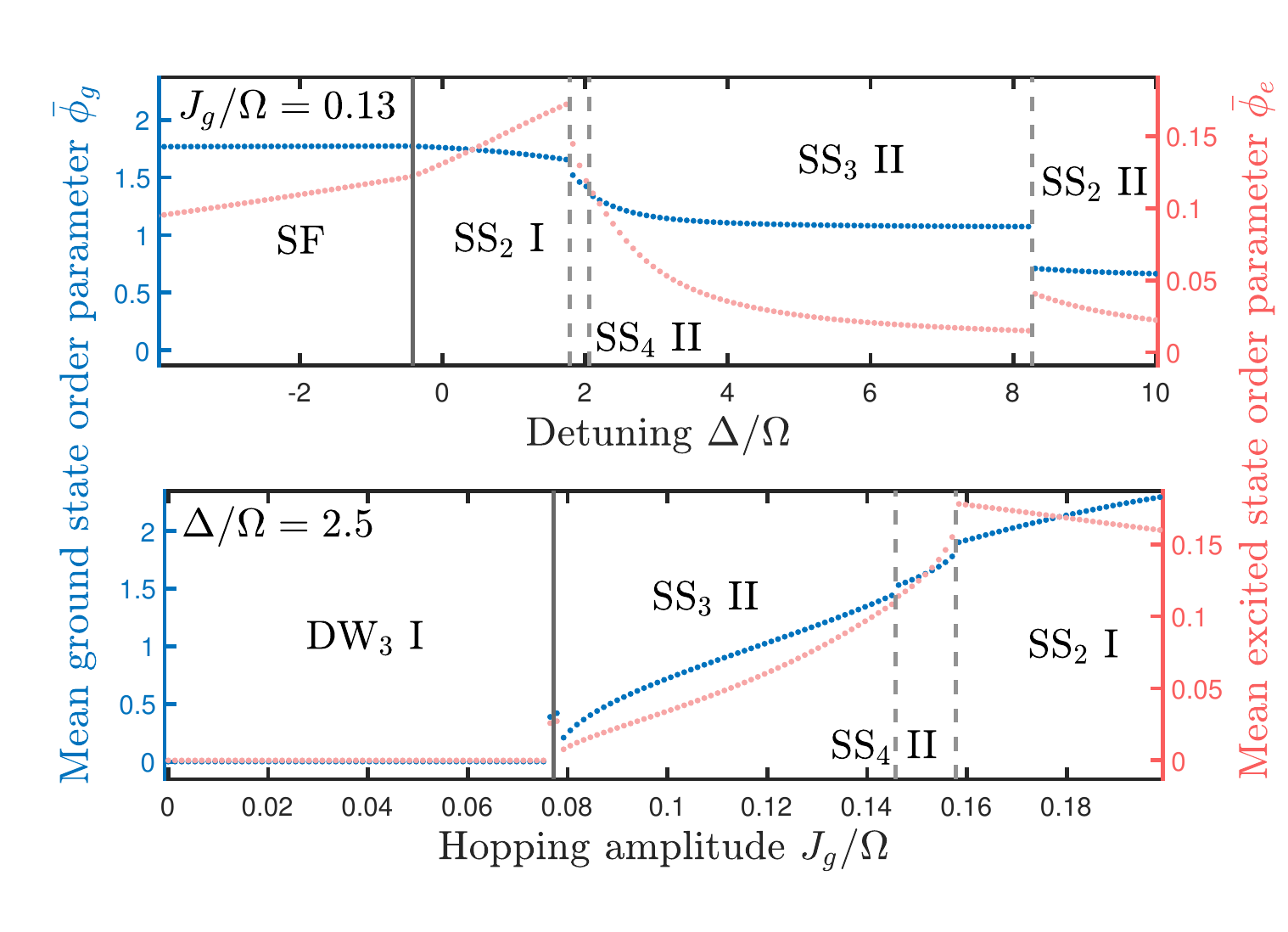}
\caption{Mean ground state order parameter $\bar \phi_g$ (dark blue line) and excited state order parameter $\bar \phi_e$ (light red line) at fixed detuning (upper diagram) and fixed hopping amplitude (lower diagram). First order phase transitions are identified through jumps of the mean observables (solid vertical line) and second order transitions are identified through kinks (dashed vertical line).}
\label{fig:phase_transitions}
\end{figure}

\subsubsection{Beyond frozen limit $J_g > 0$}
In this section we investigate the phase diagram beyond the frozen limit ($J_g > 0$), which extends the devil's staircase (see FIG. \ref{fig:anisotropic}). We obtain a DW, a SS, a SF and a vacuum regime of phases with various crystalline structure. While the DW regime only possesses phases of type I due to the negative chemical potential, we find SS phases of type I and II. For large, negative detunings, the occupation number of the excited state vanishes which implies that translational symmetry is not broken. While for small hopping amplitudes the ground state state is the vacuum state, large hopping rates lead to SF phases. For positive detunings, the occupation number of the excited state becomes finite, which results in inhomogeneous ground state phases. The DW regime corresponds to the devil's staircase and melts into a SS regime for large enough hopping amplitude. Similar to the devil's staircase, the SS regime possesses a manifold of type II phases with different crystalline structure. The SF regime and the SS II phases are separated by a SS I phase, which consists of a quantum phase featuring stripes of excited particles separated by a stripe where only the finite ground state occupation number is finite. The phase transitions from SF to SS, as well as the phase transitions within the SS regime all shift to larger detunings as the hopping amplitude increases. The phase diagram is consistent with the previously observed two-stage melting of solid phases in Rydberg-excited systems \cite{MeltingI,MeltingII,MeltingIII,MeltingIV}. An initial DW phase obtained in the frozen limit goes through two stages of melting upon increasing the hopping amplitude. The first melting consists of a transition from DW to SS, which causes only a fraction of the particle to delocalize and hereby maintains the initial crystalline structure. The second melting corresponds to a transition from SS to SF, where the increasing hopping amplitude leads to a completely delocalized homogeneous state.\\
We obtain both first-order and second-order phase transitions within the phase diagram (see FIG. \ref{fig:phase_transitions}). The two phase transitions associated with the melting process, namely the DW-SS and the SS-SF transitions, are second-order, which is consistent with results obtained in the previous works. Interestingly, we find that the system undergoes first-order phase transitions between both melting stages within the SS regime. Since the phases within this regime possess different superlattice areas with discrete values, the transition from one SS phase to another leads to a jump of the spatially averaged observables. We identify these jumps as first-order phase transitions between the various SS phases when either the detuning or the hopping amplitude is increased.\\
Phase diagrams with similar regimes have been obtained in previous calculations for Rydberg-excited bosons with isotropic long-range interaction trapped in square and triangular optical lattices \cite{RydbergAndreas,RydbergPhasediagram2}, though the emerging crystalline structures highly depend on the lattice and interaction geometries.\\

\begin{figure}[t]
\includegraphics[width = 0.95\linewidth, trim = {10 0 0 0}, clip]{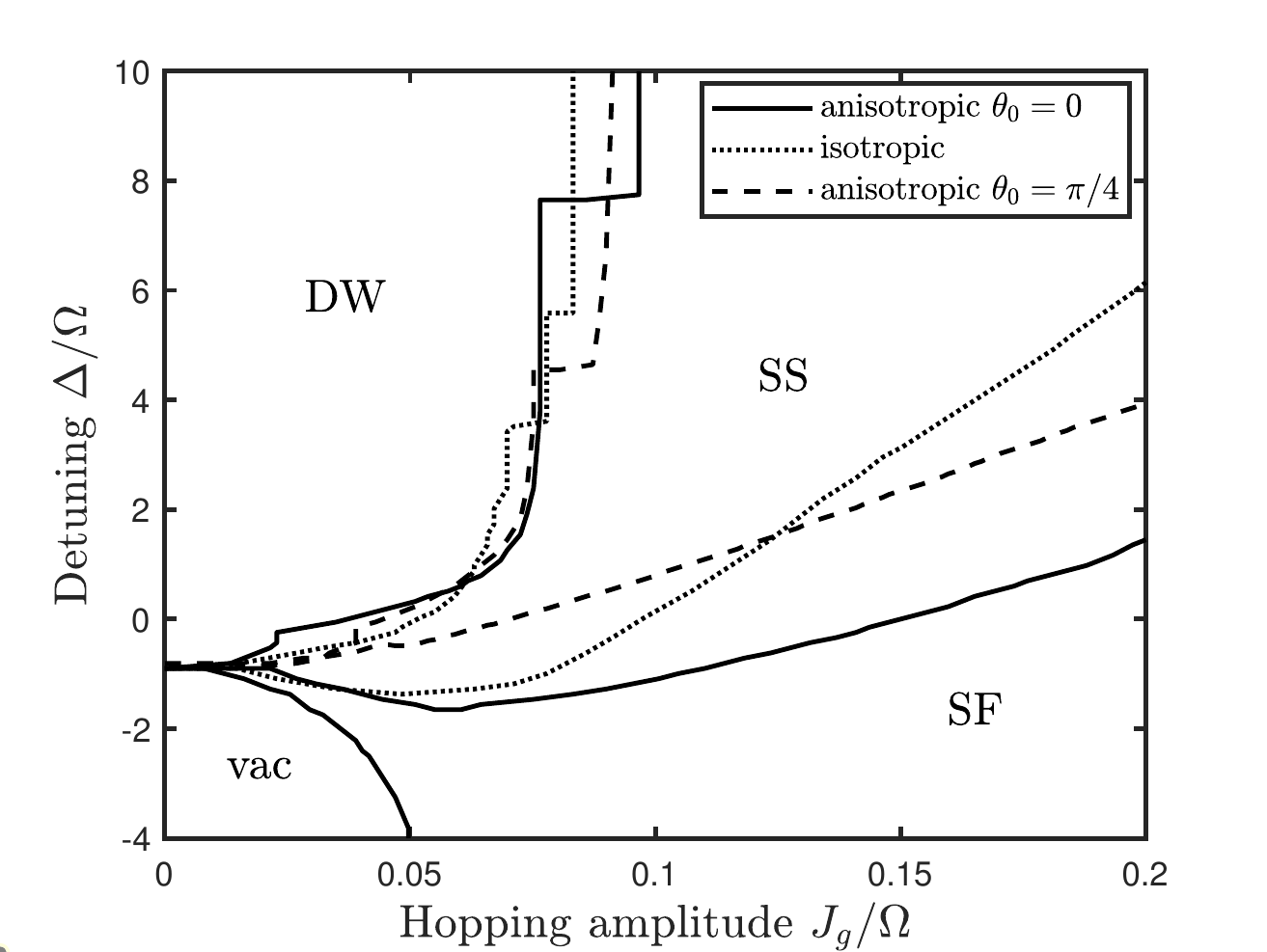}
\caption{Phase boundaries between the various regimes obtained for anisotropic (solid line), isotropic (dotted line) and tilted (dashed line) anisotropic long-range interaction at $U_g/\Omega = 0.1$, $U_e/\Omega = 1$, $U_g/\Omega = 1000$, $J_e/\Omega = 0$ and $\mu/\Omega = -0.25$. While the DW-SS phase transition does not shift significantly, we see a more pronounced difference between the three types of long-range interaction regarding the SS-SF phase transition. For a discussion of the full phase diagrams in the isotropic and tilted anisotropic cases see Appendix D.}
\label{fig:comparison}
\end{figure}

\jump
In the following, we further investigate the role of the reference axis and the interaction geometry. We therefore perform calculations with tilted anisotropic interaction ($\theta_0 = \pi/4$) and isotropic interaction ($V_0/\Omega = 0$, $V_1/\Omega = 1000$).\\
\noindent
The phase diagrams obtained with tilted anisotropic interaction and isotropic interaction exhibit features similar to the phase diagram obtained with anisotropic interaction (see FIG. \ref{fig:comparison}). Both phase diagrams show evidence of a two-stage melting process and although the DW-SS boundary appears to be only slightly affected, we find a substantial of the choice of the interaction geometry on the location of the SF-SS boundary.\\
When the reference angle is set to $\theta_0 = \pi/4$, the long-range interaction strength is maximum along one diagonal, which leads to striped phases along the perpendicular diagonal direction (see figure in Appendix D). The increased distance between two sites within a stripe impedes coherent hopping, which renders the SF energetically more favorable and thus shifts the SF-SS boundary to larger detunings for increased hopping amplitude.\\
An isotropic interaction, i.e. $V_1/\Omega = 1000$ and $V_0/\Omega = 0$, leads to unique crystalline structures composed of equidistant excited particles (see figure in Appendix D). Since the SS phases closest to the SS-SF boundary are checkerboard-ordered, which is not as favorable as stripes with respect to coherence, the boundary is shifted as well.\\
We conclude that striped phases along one coordinate axis of the square lattice are most favorable for the coexistence of finite condensation and crystalline ordering.\\

\section{Conclusion}
We computed the ground state phase diagrams of the extended, two-component Bose-Hubbard model with anistropic, tilted anisotropic and isotropic long-range interaction, and found SF, DW and SS phases as ground states for the appropriate choice of parameters. The emerging crystalline structure heavily depends on the type of interaction. We also observed that the transition between homogeneous phases and phases with broken translational symmetry shifts to larger detunings as the hopping amplitude increases. By comparison of the phase diagrams, we notice this shift to be largest for isotropic interaction. We attribute this to the geometry of the crystalline structure, since striped phases, obtained for anisotropic interaction, are more advantageous than checkerboard-ordered phases for the coexistence of coherent tunneling (leading to a condensate) and spontaneously broken translational symmetry.\\
We believe that the advantages of anisotropic interaction highlighted in this work, combined with the additional advantages of the single-photon Rydberg excitation scheme, make the dressing with Rydberg $p$-states a promising direction for further experimental research on Rydberg-excited quantum gases in optical lattices.

\section*{Acknowledgments}
We thank Alexander Kondratiev, Asok Rafael Thun, Nicolas Christophel and Georg Wille for technical support and fruitful discussions. Support by the Deutsche Forschungsgemeinschaft via DFG SPP 1929 GiRyd, DFG HO 2407/8-1 and the high-performance computing center \textit{Center for Scientific Computing} is gratefully acknowledged.

\bibliographystyle{apsrev4-1}
\bibliography{paper_reflist}

\section*{Supplementary material}
In the following we discuss technical details of the Gutzwiller mean-field theory, the Hamiltonian used for the benchmark of the Hartree approximation, the observables calculated for the identification of regimes and phase transitions, and the phase diagrams corresponding to the tilted and the isotropic interaction.

\begin{figure}[b]
\includegraphics[width = 1\linewidth, trim = {0 20 0 0}, clip]{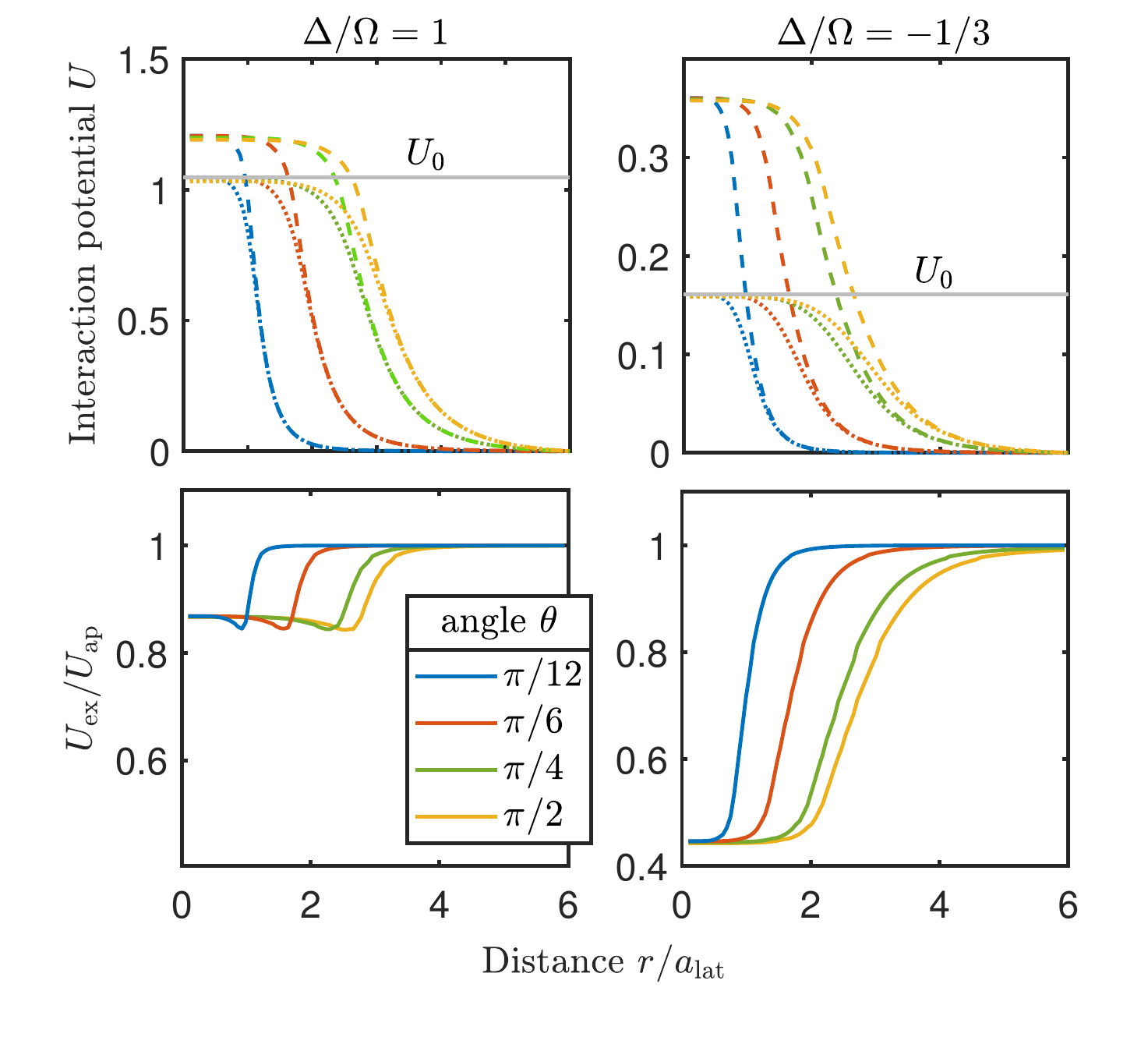}
\caption{Upper figures: Soft-core potential $U_\mathrm{ex}$ obtained by exact diagonalization of the two-body Hamiltonian (dotted line) and the soft-core potential $U_\mathrm{ap}$ within the Hartree-approximation (dashed line) for different angle $\theta$ at $V_0/\Omega = 1000$. Lower figures: Ratio $U_\mathrm{ap}/U_\mathrm{ex}$ between the exact and the approximated soft-core potentials. The approximated soft-core potentials obtained are larger than the ones obtained with the exact Hamiltonian. While the difference is small for positive detunings ($\Delta/\Omega = 1$), it becomes more important in the regime of negative detunings ($\Delta/\Omega = -1/3$). The angle affects the characteristic range $r_c$, but does not influence the quality of the approximation.}
\label{fig:softcore}
\end{figure}

\begin{figure}[t]
\includegraphics[width = 0.45\linewidth]{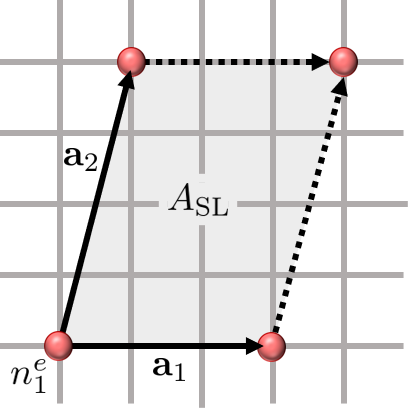}
\caption{Definition of the underlying superlattice of the crystalline structure. The superlattice is defined by two spanning vectors $\textbf{a}_1$ and $\textbf{a}_2$. As a measure for identifying quantum phases and} differentiating crystalline structures, we determine the superlattice unit cell area $A_\text{SL} = |\textbf{a}_1 \times \textbf{a}_2|$. In the non-tilted anisotropic case the superlattice unit cell area quantifies the distance between Rydberg-excited stripes and in the isotropic case the distance between Rydberg-excited particles. We also define the excited state fraction $n_1^e$ of the single occupied site in a superlattice unit cell.
\label{fig:superlattice}
\end{figure}

\begin{figure*}[t]
\includegraphics[width = 0.97\textwidth, trim = {00 0 20 0}, clip]{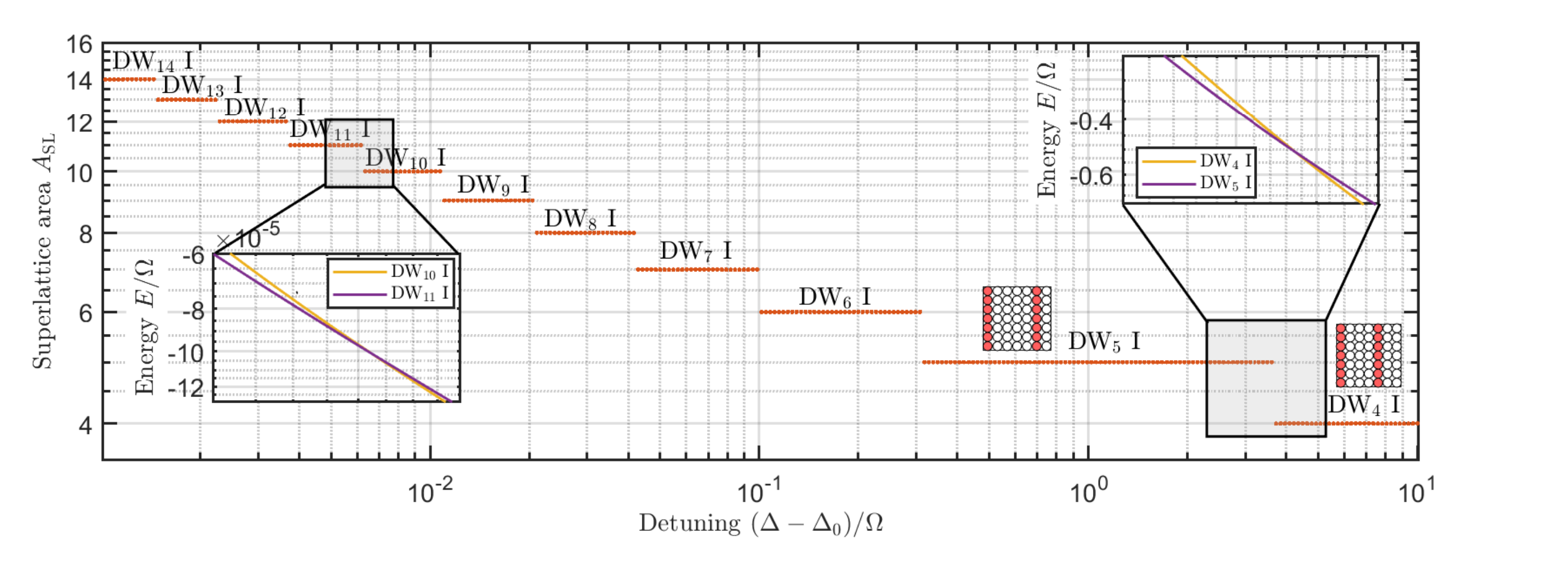}
\caption{Devil's staircase of crystalline density wave phases obtained by variation of the detuning $\Delta$ at $U_g/\Omega = 0.1$, $U_e/\Omega = 1$, $U_g/\Omega = 1000$, $V_0/\Omega = 1000$, $V_1/\Omega = 1$, $J_e/\Omega = 0$ and $\mu/\Omega = -0.25$. Each step corresponds to a density wave phase, which has the lowest energy for the chosen parameters. Insets: Energy $E/\Omega$ of the DW$_{4(10)}$ I and DW$_{5(11)}$ I by variation of the detuning $\Delta$. The energies of both density wave phases cross and the superlattice area changes across the phase transition.}
\label{fig:devil_energy}
\end{figure*}

\subsection{Mean-fields in the Gutzwiller theory}
In our model different lattice sites are coupled through the ground state hopping, the excited state hopping and the long-range interaction. Within Gutzwiller mean-field theory, we treat the full system as a set of individual lattice sites coupled to self-consistent mean-fields instead of other lattice sites. We achieve this by expanding the operators involved in the before mentioned processes in terms of their quantum fluctuations. For the hopping processes, we expand $\hat b_i = \langle \hat{b_i} \rangle + \delta \hat{b_i}$ and rewrite the hopping term as
\begin{equation}
\begin{split}
\cre_i \an_j &= \cre_i \ex{\an_j} + \an_j \ex{\cre_i} - \ex{\an_i} \ex{\cre_j} + \delta \cre_j \delta \an_i\\
&\approx \cre_i \ex{\an_j} + \an_j \ex{\cre_i} - \ex{\an_i} \ex{\cre_j},
\end{split}
\end{equation}
neglecting terms, which are quadratic in the quantum fluctuations $\delta \an_j$ and $\delta \cre_j$. By taking into account the sums of the full extended Bose-Hubbard model (see main text), we rearrange the terms
\begin{equation}
\begin{split}
&\sum_{\NN{ij}} (\cre_i \an_j  + \mathrm{h.c.}) = \sum_{\NN{ij}} (2\cre_i \ex{\an_j} - \ex{\an_i} \ex{\cre_j} + \mathrm{h.c.})\\
=& \sum_i (\cre_i \sum_{j \in \mathrm{NN}(i)} \ex{\an_j} -  \frac{1}{2}\ex{\cre_i} \sum_{j \in \mathrm{NN}(i)} \ex{\an_j}  + \mathrm{h.c.}),
\end{split}
\end{equation}
where we split the sum over $i$ and $j$ as $\sum_{\NN{ij}} = \frac{1}{2} \sum_i \sum_{j \in \mathrm{NN}(i)}$ and $\mathrm{NN}(i)$ denotes the nearest neighbor of site $i$. Even though the hopping term now depends on the neighboring observables $\ex{\an_j}$, it has become a local term and its strength is proportional to the mean-field $\xi^\nu_i = \sum_{j \in \mathrm{NN}(i)} \langle \hat{b}_i^\nu \rangle$ with $\nu \in \{g,e\}$. The second term within the sum, a constant energy offset given by the mean-fields, reads $E_i^\text{hop} = \frac{1}{2} \sum_\nu (J_\nu \ex{(b^\nu_i)^\dag} \xi^\nu_i + \mathrm{h.c.})$.\\
The long-range interaction term is decoupled in a similar fashion. We rewrite the occupation number operator using the quantum fluctuation $\delta \n_i = \hat n_i - \langle \hat n_i \rangle$ and use the Hartree approximation as
\begin{equation}
\begin{split}
\n_i \n_j &= \n_i \ex{\n_j} + \n_j \ex{\n_i} - \ex{\n_i} \ex{\n_j} + \delta \n_j \delta \n_i\\
&\approx \n_i \ex{\n_j} + \n_j \ex{\n_i} - \ex{\n_i} \ex{\n_j}.
\end{split}
\end{equation}
Within the full Hamiltonian (see main text), the long-range interaction term can then be reorganized as
\begin{equation}
\begin{split}
&\sum_i \sum_{j \neq i} V_{ij} \hat{n}^e_i\hat{n}^e_j = \sum_i (2 \hat{n}^e_i \sum_{j \neq i} V_{ij} \ex{\n^e_j} - \ex{\n^e_i} \sum_{j \neq i} V_{ij} \ex{\n^e_j})
\end{split}
\end{equation}
where we define $V_{ij} = (V_0  \mathrm{sin}^4(\theta - \theta_0) + V_1)/|\vec{r}_i-\vec{r}_j|^6$ for the sake of simplicity. The long-range interaction term is now decoupled and effectively resembles a local potential given by the mean-field $\eta_i = 2 \sum_{j \neq i} V_{ij} \langle \hat{n}^e_j \rangle$. As for the hopping term, we obtain a constant energy shift $E_i^\text{int} = - \frac{1}{2} \ex{\n^e_i}\eta_i$.\\
Finally, we combine both energy shifts to $E_i^\text{off} = E_i^\text{hop} + E_i^\text{int}$. This constant offset does not affect the self-consistency procedure of the numerical simulation, but is calculated after the ground state is reached in order to obtain the full ground state energy.\\

\subsection{Hartree benchmark calculation}
Within the RWA, the energy eigenstates of two atoms in the electronic ground state coherently coupled to an electronic excited state with Rabi coupling $\Omega$ and separated by distance $r$ are given in the two-body basis $\{|gg\rangle,|ge\rangle,|eg\rangle,|ee\rangle\}$ by the Hamiltonian
\begin{equation}
\hat{H} =
\begin{pmatrix}
0 & \Omega/2 & \Omega/2 & 0\\
\Omega/2 & -\Delta & 0 & \Omega/2\\
\Omega/2 & 0 & -\Delta & \Omega/2\\
0 & \Omega/2 & \Omega/2 & -2\Delta + V(r,\theta)
\end{pmatrix}
\end{equation}
with the van-der-Waals interaction strength $V(r,\theta) = V_0  \mathrm{sin}^4 (\theta)/r^6$ between atoms $1$ and $2$. We omit the residual isotropic part of the interaction for the sake of simplicity. Through analytic derivation of the ground state energy within perturbation theory, we obtain an effective soft-core interaction potential with height $U_0 = (\Delta - \sqrt{\Delta^2 + 2\Omega^2})/2 + \sqrt{\Delta^2 + \Omega^2}$ and characteristic range of the interaction $r_c(\theta) = [(V_0 \mathrm{sin}^4 (\theta))/(2|\Delta|)]^{1/6}$ \cite{SinglePhotonII}. Numerical computation of the ground state energy through exact diagonalization yields an almost identical soft-core potential.\\
Within the Hartree approximation, the self-consistency condition makes the derivation of an analytic expression for the soft-core potential impossible. Thus we determine the system's ground state energy numerically. After decoupling the long-range interaction term, we obtain a separate Hamiltonian for each particle in the respective $\{|g\rangle,|e\rangle\}$-basis. For atom $i \in \{1,2\}$ it is given by
\begin{equation}
\hat{H}^i_\mathrm{Hrt} =
\begin{pmatrix}
0 & \Omega/2\\
\Omega/2 & -\Delta + V(r,\theta) n_j^e 
\end{pmatrix}
+ E(n^e_j)
\end{equation}

\noindent
with the excited state occupation number $n_j^e = \langle \hat{n}_j^e\rangle$ of the other atom $j$ and the energy offset $E(n^e_j)$ resulting from the Hartree-approximation. Both effective single-atom Hamiltonians depend on the energy eigenstates of the respective other atom via the occupation number $n_j^e$, which renders the method self-consistent.\\
We determine the soft-core potential for various values of the detuning $\Delta/\Omega$ and the angle $\theta$, while keeping the interaction strength $V_0/\Omega = 1000$ fixed (see FIG \ref{fig:softcore}). The soft-core height obtained by exact diagonalization the two-body Hamiltonian fits with the height $U_0$ obtained through analytic calculation. Although the soft-core potential depends on the angle $\theta$, we find that varying $\theta$ only changes its characteristic range $r_c$, which is in agreement with its analytic expression. We therefore conclude that the quality of the Hartree-approximation only depends on the detuning $\Delta/\Omega$ (see FIG \ref{fig:softcore}).

\subsection{Identifying phase regimes and transitions}
After application of Gutzwiller mean-field theory, the Hamiltonian of each lattice site can be diagonalized separately and local observables can be calculated. For lattice site $i$ with wave function $|\Psi \rangle_i$ the relevant observables are the condensate order parameter $\phi^{\nu}_i = \langle \Psi | \hat{b}^{\nu}_i | \Psi \rangle_i$ and the occupation number $n^{\nu}_i = \langle \Psi | \hat{n}^{\nu}_i | \Psi \rangle_i$ of the ground ($\nu = g$) and excited ($\nu = e$) state. Although these calculations are reduced in computational complexity compared to the original problem, finding the ground state for given parameters is no trivial task. Inhomogeneous phases, which arise due to the long-range interaction, require a superlattice description of the system \cite{RydbergAndreas,RydbergPhasediagram2}. For periodic boundary conditions, the crystalline structure of Rydberg-excited particles can be described by two a superlattice unit cell with spanning vectors $\vec{a}_1$ and $\vec{a}_2$ (see FIG. \ref{fig:superlattice}). Since the spatial distribution of Rydberg-excited particles follows the geometry of the chosen superlattice unit cell instead of spontaneously choosing a spatial ordering, it is necessary to perform the ground state calculation for different superlattices. For given parameters, the correct many-body ground state is chosen to be the ground state corresponding to the superlattice which yields the lowest energy.\\
In our calculation, we limit the choice of the spanning vectors $\vec{a}_1$ and $\vec{a}_2$ by setting a maximum value of the superlattice unit cell area $A_\mathrm{SL} = |\vec{a}_1 \times \vec{a}_2|$. We found $A_\mathrm{SL}^\mathrm{max} = 18$ to be a reasonable cutoff, as no superlattice unit cell area of the many-body ground states obtained through the computation exceeded that value. Larger superlattice unit cell areas become relevant in the devil's staircase calculation for detunings $\Delta$ closer to the critical detuning $\Delta_0$ than the values considered in this work.\\
We then compute the phase diagram by calculation of the many-body ground states for all considered superlattices and comparison of their energies (see FIG. \ref{fig:devil_energy}). In order to establish different phases in the phase diagram, we define quantities which allow us to distinguish quantum phases more easily. The superlattice unit cell area already allows us to identify dense or sparse distributions of Rydberg-excited particles in the system. The case $A_\mathrm{SL} = 1$ corresponds to a homogeneous phase. We then define spatially averaged values of different observables within a superlattice unit cell, specifically the mean condensate order parameter $\bar{\phi}_{\nu} = \sum_{i \in A_\mathrm{SL}} |\phi^{\nu}_i| / A_\mathrm{SL}$ and the mean occupation number $\bar{n}_{\nu} = \sum_{i \in A_\mathrm{SL}} |n^{\nu}_i|/A_\mathrm{SL}$ of the ground (excited) state.\\
\begin{table}[H]
\begin{center}
\begin{tabular}{| c || c c c |}
\hline
Phase & $\bar{\phi}$ & $\bar{n}$ & $A_\mathrm{SL}$ \\ 
\hline
\hline
MI & 0 &$\mathbb{N}^+$ & 1 \\ 
\hline
SF & $\mathbb{R}$ & $\mathbb{R}$ & 1 \\ 
\hline
DW & 0 & $\mathbb{Q}$ & $\mathbb{N}^+/\{1\}$ \\ 
\hline
SS & $\mathbb{R}$ & $\mathbb{R}$ & $\mathbb{N}^+/\{1\}$ \\ 
\hline
\end{tabular}
\caption{Classification of the Mott insulating (MI), the superfluid (SF), the density wave (DW) and the supersolid (SS) phase by means of the mean condensate order parameter $\bar{\phi}$, the mean occupation number $\bar{n}$ and the area $A_\mathrm{SL}$.}
\label{tab:phases}
\end{center}
\end{table}
\noindent
In table \ref{tab:phases}., we depict the classification of the various quantum phases. Differentiating between the various DW and SS phases is especially difficult, since several quantum phases in the chosen parameter space can share the same order parameters despite distinct spatial distributions. Hence we also depict the spatial distribution of the observables within the system. This also helps us identifying whether the inhomogeneous phases obtained are of type I or type II.\\
\noindent
As the phase transition is obtained through the comparison of the energies of neighbouring phases in the phase diagram, we identify whether a phase transition is of first or second order via a discontinuous or continuous change of these mean observables upon parameter change.
\begin{figure}[b]
\includegraphics[width = 1\linewidth, trim = {0 0 0 0}, clip]{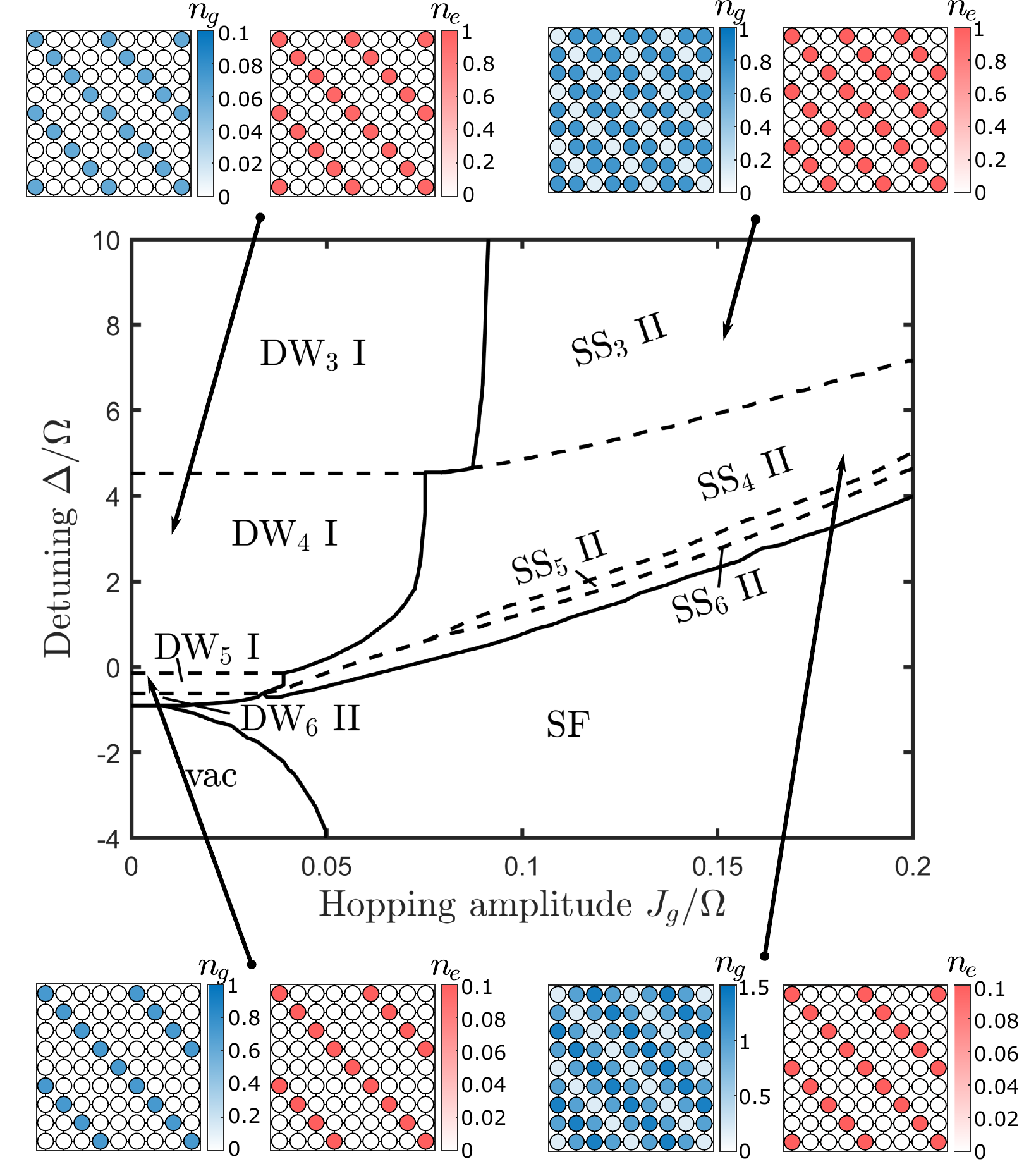}
\caption{$J_g-\Delta$ phase diagram obtained with tilted anisotropic long-range interaction ($\theta_0 = \pi/4$) at $U_g/\Omega = 0.1$, $U_e/\Omega = 1$, $U_g/\Omega = 1000$, $V_0/\Omega = 1000$, $V_1/\Omega = 1$, $J_e/\Omega = 0$ and $\mu/\Omega = -0.25$. The spatially modulated ground state phases are characterized by diagonal stripes of Rydberg-excited particles. While the phase diagram resembles the one obtained for the non-tilted anisotropic interaction without tilt (see figure in main text), it lacks a SS I phase. The indices denote the superlattice area $A_\text{SL}$.}
\label{fig:angles}
\end{figure}

\begin{figure*}[t]
\includegraphics[width = 0.97\textwidth, trim = {40 0 10 0}, clip]{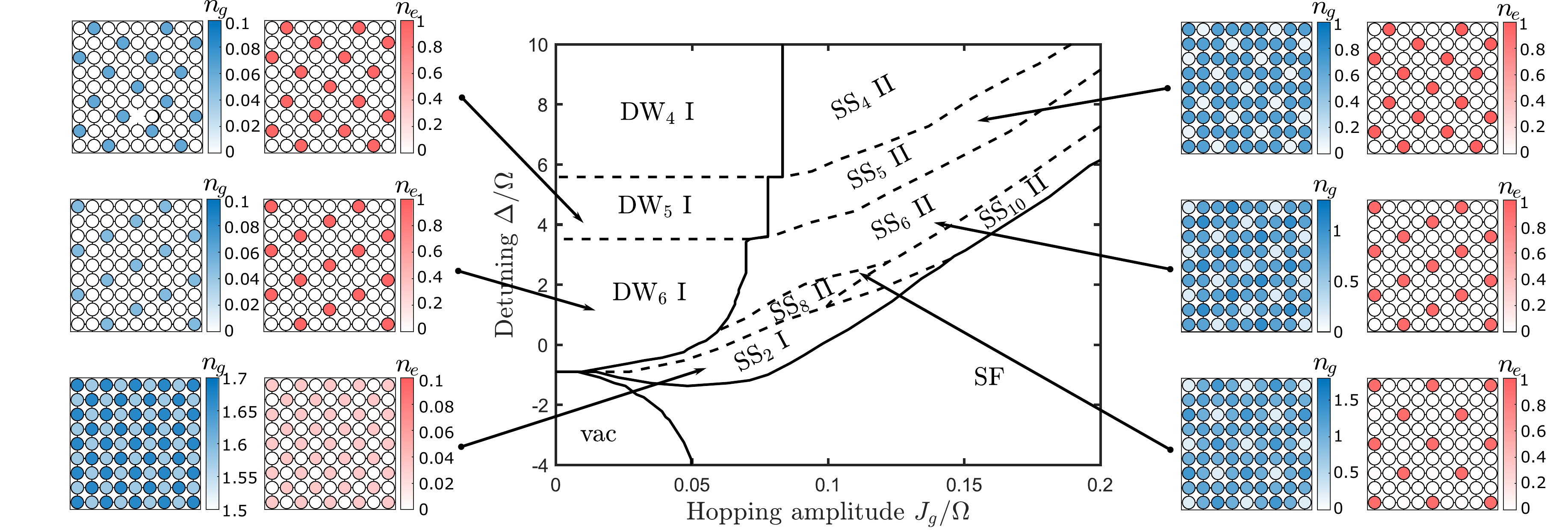}
\caption{$J_g-\Delta$ phase diagram obtained with isotropic long-range interaction at $U_g/\Omega = 0.1$, $U_e/\Omega = 1$, $U_g/\Omega = 1000$, $V_0/\Omega = 1$, $V_1/\Omega = 1000$, $J_e/\Omega = 0$ and $\mu/\Omega = -0.25$. The phase diagram exhibits various regimes with different crystalline structures composed of equidistant Rydberg-excited particles. The indices denote the superlattice area $A_\text{SL}$. These results are in agreement with the ones obtained in \cite{RydbergAndreas}.} 
\label{fig:isotropic}
\end{figure*}
\subsection{Phase diagrams in the tilted and isotropic case}
Calculation with a finite reference angle $\theta_0 \neq 0$ leads to a phase diagram similar to the $\theta_0 = 0$ case. The most noticeable difference is the narrower SS regime due to extended DW and SF regimes (see figure in main text). Since the stripes are prone to emerge perpendicular to the reference axis due to the nature of the long-range interaction, a tilted reference axis leads to tilted stripes (see FIG. \ref{fig:angles}). For the case $\theta_0 = \pi/4$ we obtain diagonal stripes connecting next-nearest neighbours in the system. The increased distance between particles reduces coherent tunneling within a stripe and therefore renders a condensate along the stripes energetically less favorable, which leads to a smaller SS regime compared to the $\theta_0 = 0$ case within the considered parameter space. This is in agreement with the numerical results obtained.\\
For isotropic interaction we find many different quantum phases with unique crystalline orders (see FIG. \ref{fig:isotropic}). The isotropic interaction leads to spatial distributions defined by equidistant Rydberg-excited particles, which yields in a square two-dimensional optical lattice interparticle distances of $d/a_\mathrm{lat} \in \{1, \sqrt{2}, 2, \sqrt{5}, \sqrt{8}, ...\}$. The anisotropic interaction, on the other hand, favors striped phases for which at $\theta_0 = 0$ typical distances are $d/a_\mathrm{lat} \in \{1, 2, 3, ...\}$ and therefore fewer different spatial distributions for equal lattice system sizes.\\
Similar to the non-tilted anisotropic case we obtain also for isotropic interaction a SS$_2$ I phase with checkerboard-like spatially modulated density, which is the smallest, inhomogeneous crystalline order possible. With respect to superfluid flow of particles, this phase is not as favorable as its SS$_2$ I striped counterpart, since it allows for coherent tunneling along one direction. The SS$_2$ I regime therefore narrows and vanishes for increasing hopping amplitude.

\end{document}